\documentclass{IEEEtran}
\usepackage{amsmath}
\usepackage{amssymb}
\usepackage{graphicx}
\usepackage{cite}
\DeclareGraphicsExtensions{.pdf,.jpg,.png,.jpeg}

\begin{document}
	
\title{
Logarithmic resilience risk metrics that address the huge variations in blackout cost
}

\author{Arslan Ahmad and Ian Dobson, Iowa State University, August 2025
\thanks{A. Ahmad and I.~Dobson are with Dept. Electrical \& Computer Engineering, Iowa State University, Ames Iowa USA; email: dobson@iastate.edu. 
Support in part from NSF grants 2153163 and 2429602, Argonne National Laboratory, US DOE EERE Award DE-EE0010724, and PSerc project S110 is gratefully acknowledged. The views expressed herein do not necessarily represent the views of US DOE or the United States Government.}}


\maketitle	

\begin{abstract}
Resilience risk metrics must address the customer cost of the largest blackouts of greatest impact. 
However, there are huge variations in blackout cost in observed distribution utility data that make it impractical to properly estimate the mean large blackout cost and the corresponding risk. 
These problems are caused by the heavy tail observed in the distribution of customer costs. 
To solve these problems, we propose resilience metrics that describe large blackout risk using the mean of the logarithm of the cost of large-cost blackouts, the slope index of the heavy tail, and the frequency of large-cost blackouts.
\end{abstract}


\section{Events, cost, and metrics}
\label{metrics}

For a resilience analysis, we extract from historical outage data recorded by utilities events in which outages bunch up and overlap \cite{CarringtonPS21,StankovicPS23}.
Alternatively, the events could be simulated with power system models \cite{PoudelSJ20}. 
The events we observe range in size from isolated single outages to large events with hundreds of outages caused by extreme weather. 
The customer costs for each event can be estimated based on factors such as the customer hours without power and the types of customers. 
Let $n_{\rm customer}$ be the number of customers served in a distribution system.
Then the 
cost of the $i$th event in the distribution system normalized by the number of customers served is
\begin{align}
c_i=\frac{\mbox{customer cost for event }i}{n_{\rm customer}}
\end{align}
Suppose there are $n$ events with normalized costs\footnote{Customer costs are assumed positive by neglecting events with zero cost.}
\begin{align}
    C_{\rm all}=\{c_1,c_2,\cdots,c_n\}.
\end{align}
For a resilience analysis we are most interested in events with large cost, so we set a threshold  $c_{\rm large}$ and define the costs of the $n_{\rm large}$ large cost events as
\begin{align}
    C_{\rm large}&=\{\mbox{costs $c$ in $C_{\rm all}$ such that $c\ge c_{\rm large}$}\}
    \label{clarge}
\end{align}
If the number of years of outage data is $n_{\rm year}$, the average annual rate of events with positive customer cost is $E_{\rm rate}=n/n_{\rm year}$.
Then we easily obtain empirical estimates of the probability  $p_{\rm large}$ of large cost events and its associated annual rate $f_{\rm large}$ and recurrence interval $\rm RI_{\rm large}$:
\begin{align}
    p_{\rm large}&=\frac{n_{\rm large}}{n};\ 
    f_{\rm large}= p_{\rm large}\, E_{\rm rate};\  {\rm RI_{\rm large}}=f_{\rm large}^{-1}
    \end{align}
\looseness=-1
For this letter we choose $c_{\rm large}$ so that $p_{\rm large}=0.1$ or so that $p_{\rm large}=0.05$. The choice of $c_{\rm large}$ is discussed at the end of section~\ref{heavytail}. 
$c_{\rm large}$ is kept fixed once chosen to allow comparisons of large event risk over time.

Now we quantify the risk of large cost events.
The risk of large events is sometimes defined as the probability of a large event multiplied by the mean cost of large events. However, the large event costs vary over several orders of magnitude. 
To address this, and to avoid the severe difficulties caused by attempting to estimate the mean cost that are discussed in Section~\ref{heavytail}, we instead transform the large cost data by applying a logarithm before taking the mean. 
That is, we define the Average Log Event Cost metric ALEC for large cost events as
\begin{align}
\mbox{ALEC}
   & =  \frac{1}{n_{\rm large}}
  \sum_{c\in C_{\rm large}}\log_{10}c
   \label{ALEClarge}
\end{align}
The use of the logarithm to base ten shows that 
ALEC indicates the mean order of magnitude of the large event costs. For example, ${\rm ALEC}=2$ indicates costs of mean order of magnitude 2, or $10^2$. Note that
$10^{\rm ALEC}$ is the geometric mean of the large event costs.

\looseness=-1
We also define the Annual Log Cost Resilience Index ALCRI for large cost events, which usefully combines frequency and ALEC, as
\begin{align}
\mbox{ALCRI}
   & =  \frac{1}{n_{\rm year}}\sum_{c\in C_{\rm large}}\log_{10}c=f_{\rm large}\,{\rm ALEC}
   \label{ALCRIlarge}
\end{align}

It is well-known in reliability that there is huge variability in blackout sizes and correspondingly large variations in the SAIDI reliability metric if extreme events are included.   
Indeed, the distribution of daily SAIDI typically follows a heavy-tailed lognormal distribution, and major event days are defined and often excluded from the calculations if their daily SAIDI is in the tail of the distribution \cite{IEEE13662022}. 
Instead of excluding the tail events as needed for consistent year-to-year estimates of reliability, here we address resilience by characterizing the tail of the distribution of event cost with metrics.

The  Area Index of Resilience AIR metric \cite{PandeyArXiv25} also averages the logarithm of customer hours to reduce its variability, and Carreras \cite{CarrerasPS16} discusses the corresponding heavy tail problem for blackout sizes in transmission systems.
In fields other than power system resilience engineering, it is usual to fit the statistics of extreme events with heavy-tailed distributions and to estimate the tail index $\alpha$ with Hill estimators \cite{ClausetSIAM09,Nairbook22,Resnickbook07}.


\section{Utility data processing}

We process the outage data of five different Northeastern and Midwestern US utilities. 
The outage data includes the start time, end time, and number of customers affected by each outage.
Only the unplanned distribution system outages are selected from the data and analyzed.
Outage events are extracted from the outage data after data cleaning (details in \cite{AhmadCIRED24}), and the cost of each event is calculated by multiplying the total customer hours lost in that event by a factor $k$, where $k$ is the average cost per customer per hour. 
$k$ accounts for the widely varying costs of residential, commercial, and industrial loads by assuming that each event has the proportion of these loads outaged that is the same as the proportion of these loads in the entire system \cite{AhmadCIRED24}. 
Table~\ref{utilitydata} gives data and metrics for each utility.
We gratefully acknowledge Muscatine Power and Water for generously providing the Utility-2 outage data.

\section{Risk and consequences of heavy tails}
\label{heavytail}

\begin{figure}[hbt]
    \centering
    \includegraphics[width=0.8\linewidth]{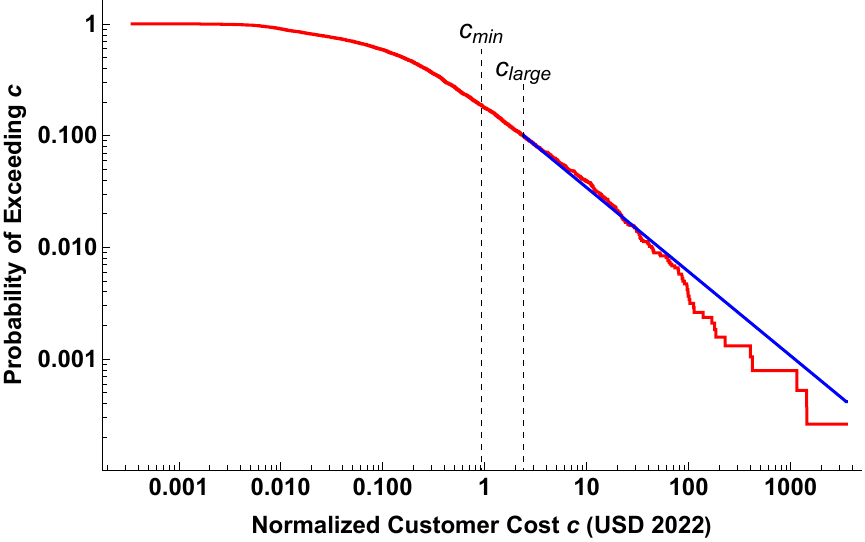}
    \caption{Exceedance curve of normalized customer cost from utility data of Utility-3, along with fitted Pareto distribution.}
    \label{fig:exceedance}
\end{figure}

\begin{figure}[hbt]
    \centering
    \includegraphics[width=0.8\linewidth]{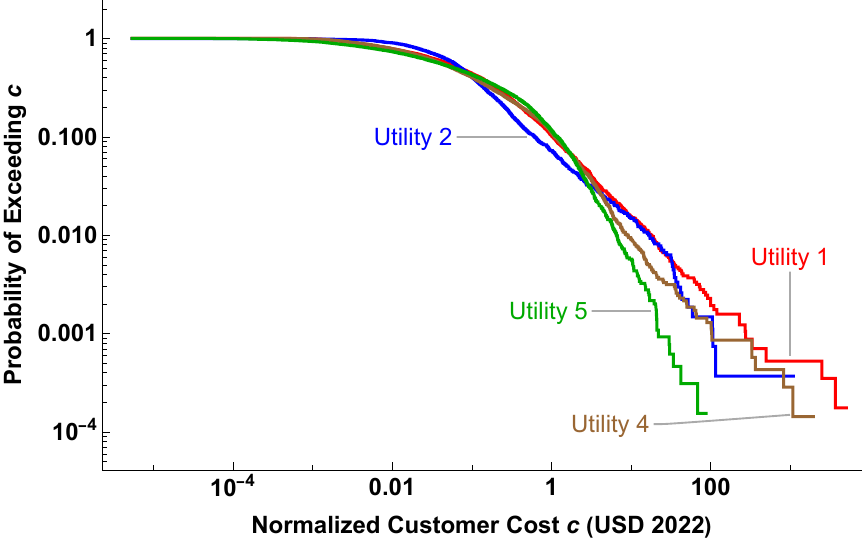}
    \caption{Exceedance curves of normalized customer cost for utilities 1,2,4,5.}
    \label{fig:combinedCCDFplot1}
\end{figure}

\looseness=-1
Risk associates probabilities with costs of events \cite{KaplanRA81}, and can be described by a probability distribution as a cost
exceedance function $\overline{F}_C(c)={\rm P}[C>c]$, which is the probability of the event cost $C$ exceeding the value $c$ as $c$ varies.\footnote{The cost exceedance function is also known as the survival function or complementary cumulative distribution
function (CCDF) or risk curve.} Fig.~\ref{fig:exceedance} shows the exceedance function $\overline{F}_C$ of per-event customer cost for Utility-3 on a log-log plot. 
Fig.~\ref{fig:exceedance}
shows an approximate straight-line behavior of slope magnitude 
$\alpha=0.75$
in the tail for $c\ge c_{\rm large}=2.4$. 
That is,  the tail in Fig.~\ref{fig:exceedance} has the
approximate power-law behavior\footnote{To verify the straight-line behavior on the log-log plot, take
the log of (\ref{powerlaw}) to obtain $\log\bar{F}(c)=-\alpha\log c + \alpha \log c_{\rm large}$} described  by a Pareto distribution $P$ with minimum value $c_{\rm large}$ and tail slope index $\alpha$:
\begin{align}
    \overline{F}_P(c)=
    (c/c_{\rm large})^{-\alpha}
    ,\qquad c\ge c_{\rm large}
    \label{powerlaw}
\end{align}
Formally testing the tails for all five utilities indicates that 
linear approximations to the tails on the log-log plots are reasonable.\footnote{
The plausibility of a Pareto (straight-line) tail is not rejected for utilities 1,2, and 5 at p=0.1, using Clauset's goodness-of-fit test \cite{ClausetSIAM09},
and the plausibility of a truncated lognormal fit is not rejected at p=0.1 for utilities 1,2,3, and 4. 
The truncated lognormal fit is quadratic, but with a large enough standard deviation parameter $\sigma$ so that the quadratic coefficient $-1/(2\sigma^2)$  is small \cite[section 1.2.3]{Nairbook22} and the linear approximation is acceptable.
A null hypothesis $H_0$ of Pareto fit above $c_{\rm large}$ is not rejected at p=0.1 for utilities 1,2,5, and $H_0$ of truncated lognormal fit above $c_{\rm large}$ is not rejected at p=0.1 for all utilities.
A likelihood ratio test between Pareto and lognormal above $c_{\rm large}$ shows insufficient evidence to prefer one over the other at p=0.1.
} 
Fig.~\ref{fig:combinedCCDFplot1} shows the empirical exceedance curves for utilities 1,2,4,5.

If $\alpha\le 1$, the mean of Pareto distribution (\ref{powerlaw}) is infinite. 
If $1<\alpha\le 2$,  the mean is $(\alpha~c_{\rm large})/(\alpha-1)$, but the standard deviation  is infinite. 
The relative standard error of an estimate of the mean with $M$ samples is infinite for $\alpha\le2$. This means that even when $1< \alpha\le2$, the mean large blackout cost is hard to accurately estimate.
Table \ref{utilitydata} shows that $\alpha<2$ for all the utilities analyzed in this letter.

In contrast, the logarithm of the data in the Pareto tail $\log_{10}P$ for  $c \geq c_{\rm large}$ has a shifted-exponential distribution with  rate parameter $\alpha\, \rm{ln}10$ and  shift parameter $\log_{10}c_{\rm large}$:
\begin{align}
\overline{F}_{\rm exp}(x)
    &=e^{-(\alpha \ln\!10) (x-\log_{10}c_{\rm large})},\quad x\ge \log_{10}c_{\rm large}
    \label{exp}
\end{align} 
If the cost data follows the Pareto tail (\ref{powerlaw}) for $c \geq c_{\rm large}$, then ALEC $=\log_{10}c_{\rm large}+(\alpha \ln\!10)^{-1}$ is the mean of the
log transformed data (\ref{exp}).
The relative standard error of an estimate of ALEC with $M$ samples is then
${\rm RSE_{ALEC}}=[(1+\alpha \ln c_{\rm large})\sqrt{M}]^{-1}$.
While the heavy Pareto tail (\ref{powerlaw}) has an infinite mean, the mean (ALEC) of the log-transformed tail (\ref{exp}) follows the usual statistics of light-tailed distributions for all values of $\alpha>0$.

The tail index $\alpha$ is the absolute slope of the tail above $c_{\rm large}$ on the log-log plot of the exceedance curve, so that $\alpha$ describes a linear trend for the observed risk\footnote{A {\sl larger} absolute slope $\alpha$ {\sl reduces} large blackouts. $\alpha$ does not depend on the multiplicative scaling of the costs or $k$ or $n_{\rm customers}$.}.  $\alpha$ can be estimated  using the maximum likelihood Hill estimator \cite{Nairbook22,Resnickbook07}, and related to ALEC according to: 
\begin{align}
\alpha&= \Big[\frac{1}{n_{\rm large}}\sum_{c\in C_{\rm large}}\hspace{-2mm}\ln\frac{c}{c_{\rm large}}\Big]^{-1}\hspace{-3mm}
=\frac{(\ln\!10)^{-1}}{{\rm ALEC-\log_{10}} c_{\rm large}}
\label{alpha}
\end{align}

The probability $p_{\rm large}$ and the absolute slope $\alpha$ describe the large cost portion of the exceedance curve \cite{AhmadCIRED24},
so this is also the case for $p_{\rm large}$ and $\rm ALEC$.
That is, $p_{\rm large}$ and $\alpha$ describe the trend for the large blackouts, and reducing $p_{\rm large}$ or $\rm ALEC$ improves this trend and reduces the risk of blackouts. 
If the large blackout cost trend in Fig.~\ref{fig:exceedance} is quite reasonably extrapolated linearly beyond the maximum observed cost (as explained in the next section), then reducing $p_{\rm large}$ or $\rm ALEC$ also reduces the risk of large blackouts beyond the largest observed.

Now we discuss choosing the 
$c_{\rm large}$ threshold for large cost events.
First, $c_{\rm large}$ should be chosen in the approximately linear upper region of the log-log plot of the exceedance curve. The linear region 
can be determined by $c \geq c_{\rm min}$, where $c_{\rm min}$ is estimated using Clauset's method \cite{ClausetSIAM09}.
Second, there is a tradeoff:   a smaller $c_{\rm large}$ includes more large event data, giving a less variable estimate of $\alpha$ and ALEC, whereas a larger $c_{\rm large}$ better describes the trend indicated by the very largest cost events,
especially if there could be some change in slope for the very largest cost events.
For this letter, we choose $c_{\rm large}$ corresponding to $p_{\rm large}=$ 0.1 or 0.05 with the result that $c_{\rm large}$ is above but near $c_{\rm min}$ to minimize the variability of the estimates of $\alpha$ and ALEC.
Determining a suitable value of $c_{\rm large}$, or, more precisely, the percentile corresponding to $c_{\rm large}$, is a well-known delicate issue in estimating $\alpha$ with the Hill estimator (\ref{alpha}) \cite{Nairbook22,Resnickbook07}. 

\begin{table}[hbpt]
	\caption{Utility data and metrics} 
	\label{utilitydata}
	\centering
\begin{tabular}{ l @{\hspace{-1pt}} c c c c c}
                        &Utility-1  &Utility-2  &Utility-3  &Utility-4  &Utility-5    \\ \hline
$n$                     & 5716      & 2706      & 3830      & 7000      & 6485     \\
$n_{\rm year}$          & 6         & 17.4      & 11        & 10        & 11     \\
$k$                     & 370.2     & 228.2     & 323.0     & 339.9     & 339.9     \\
$c_{\rm maxobs}$        & \$5063    & \$1095    & \$3523    & \$1954    & \$88     \\[1mm]

$c_{\rm large}$         & \$1.05    & \$0.62    & \$2.40    & \$1.10    & \$2.15     \\
$n_{\rm large}$         & 572       & 271       & 384       & 701       & 325     \\
$p_{\rm large}$         & 0.1       & 0.1       & 0.1       & 0.1       & 0.05     \\
$f_{\rm large}$         & 96        & 16        & 35        & 70        & 30     \\[1mm]
$\alpha$                & 0.83      & 0.73      & 0.75      & 1.00      & 1.47     \\
$\rm CI_\alpha$         & (.76,.89) & (.65,.82) & (.68,.83) & (.93,1.1) & (1.3,1.6)     \\[1mm]

ALEC                    & 0.55      & 0.38      & 0.96      & 0.48      & 0.63     \\
${\rm CI}_{\rm ALEC}$   & (.50,.59) & (.31,.46) & (.90,1.0) & (.44,.51) & (.60,.66)     \\
RSE$_{\rm ALEC}$        & 0.040     & 0.094     & 0.031     & 0.034     & 0.026     \\
RSE$_{\rm Pb}$          & 1.44      & 1.70      & 1.05      & 1.46      & 0.66      \\
RSE$_{\rm LNb}$         & 1.38      & 1.53      & 0.65      & 0.89      & 0.13      \\
\hline\\[-2.2mm]
\multicolumn{4}{l}{}
{\footnotesize all costs in 2022 USD; $f_{\rm large}$ in per year.}
\end{tabular}
\end{table}

\section{Extrapolating to largest possible cost}
\label{extrapolate}
\looseness=-1
We extrapolate the slope of the exceedance curve beyond the maximum observed cost $c_{\rm maxobs}$ so that the likely performance of the metrics can be evaluated while including the unobserved future blackouts of highest risk.  
The extrapolation of the trend is somewhat uncertain, but the extrapolation nevertheless seems more reasonable than ignoring the even more extreme events indicated by the observed trend---it seems unlikely that the largest cost blackout has already been observed in the data. 
On the other hand, in (\ref{powerlaw}) the cost has no upper bound, and that is also unrealistic. 
This section considers two ways to extrapolate the observed data up to a maximum possible blackout cost.

There is some uncertainty about the maximum possible blackout cost. To obtain a rough estimate, we suppose that the maximum possible blackout cost $c_{\rm max} = 744\, k$ corresponds to the customer cost of a total blackout lasting one month.

To extrapolate the exceedance curve tail beyond $c_{\rm maxobs}$, we fit two different distributions to the observed costs exceeding $c_{\rm large}$. Both distributions have large but finite mean and variance.
A variant  $P_b$ of the Pareto distribution (\ref{powerlaw}) with upper bound $c_{\rm max}$ is given by
$\overline{F}_{\!P_b}(c)=
(c^{-\alpha} -c_{\rm max}^{-\alpha})/(
c_{\rm large}^{-\alpha}-c_{\rm max}^{-\alpha})$ for $ c_{\rm large}< c\le c_{\rm max}$.
A bounded variant $LN_{b}$ of the lognormal distribution conditions the standard lognormal distribution 
to have $c_{\rm large} < c\leq c_{\rm max}$. 
We calculate the relative standard error of 
estimates of the mean of $P_{b}$ and $LN_{b}$ with $M$ samples as RSE = (standard deviation)/((mean)$\sqrt{M}$).

We also extrapolate to estimate the variability of ALEC:
Assuming the Pareto tail (\ref{powerlaw}), which extrapolates beyond $c_{\rm maxobs}$,
ALEC is the mean of (\ref{exp}), and we calculate the relative standard error of estimates of ALEC, $\rm RSE_{ALEC}$, using (\ref{exp}).\footnote{Upper bounding (\ref{exp}) with $c_{\rm max}$ makes no practical difference.}

\section{Results}
\looseness=-1
Table~\ref{utilitydata} shows all the results.
The empirical customer cost exceedance curves have an approximately linear trend for the larger cost events on a log-log plot. In particular, utilities  1,2,4 have  very heavy tails of absolute slope  $\alpha< 1$, utility 3 has   $\alpha\approx 1$, and utility 5 has a heavy tail with
$1<\alpha<2$. 
These heavy tails indicate that large cost events show high risk and great variability, and that there is no typical large cost event. 
The less severe and more resilient case of utility 5 can likely be attributed to it being primarily urban.
ALEC and the associated slope index $\alpha$ can be practically estimated with reasonable variability, as shown by their
95\% confidence intervals $\rm CI_{ALEC}$ and $\rm CI_{\alpha}$ for events observed over periods from 6 to 17 years.

\looseness=-1
However, there are severe difficulties in estimating the mean costs of the large events without taking the logarithm.
We extrapolated the exceedance curve beyond the maximum observed cost in two ways ($P_b$ and $LN_b$) to 
a maximum possible blackout.  
This gives large, finite means, but with   high variability in estimates of the means: The relative standard errors $\rm RSE_{Pb}$ and $\rm RSE_{LNb}$ 
are at least an order of magnitude larger than $\rm RSE_{ALEC}$ for utilities 1,2,3,4, and at least 5 times larger than 
$\rm RSE_{ALEC}$ for utility 5.
Therefore, utilities 1,2,3,4 require at least two orders of magnitude more observed events to achieve the same variability of estimates of the mean as ALEC, showing that estimating the means of the untransformed data with reasonable accuracy requires impractically large amounts of data.
Utility 5 requires an order of magnitude more observed events.

\section{Conclusions}
\looseness=-1

\looseness=-1
We quantify risk with a customer cost exceedance curve for five distribution utilities by processing their detailed outage data into events and assuming that the event customer cost is proportional to the customer hours lost. 
The empirical customer cost exceedance curves have heavy tails, implying that the risk of large-cost events is very high despite their rarity. 
It is impractical to estimate the mean cost of large events for our utility data, because the available amount of data is insufficient 
to get a workable estimate of the mean. Moreover, there is no representative or typical large-cost event.
These conclusions 
imply that the usual practices of computing metrics involving or optimizing means for extreme event risk can be invalid. 
For example, Value at Risk remains valid \cite{PoudelSJ20}, but Conditional Value at Risk, which involves taking a mean value, cannot be applied.
While we are optimistic that the heavy tails will be observed more universally so that these conclusions apply more broadly, the heaviness of the tails in the data of other utilities must be individually checked.

\looseness=-1
However, we do suggest metrics that can describe the extreme event risk by characterizing the tail of the cost exceedance curve with the probability $p_{\rm large}$ of exceeding a large event cost and either the Average Log Event Cost ALEC metric or the slope index $\alpha$. 
$\alpha$ can be calculated with a Hill estimator, and ALEC is directly related to $\alpha$. $p_{\rm large}$ can also be converted to the annual frequency of large cost events $f_{\rm large}$, and $f_{\rm large}$ can be multiplied by ALEC to yield the 
Annual Log Cost Resilience Index ALCRI.
These metrics should be used instead of relying on mean values of extreme event cost to get practical and meaningful results.
ALEC is easy to explain and calculate: take the average of the logarithms of the costs of the large cost events. 
Similarly, ALCRI is an annual sum of the logarithms of the costs of the large cost events.
ALEC and ALCRI work since taking the logarithm of heavy-tailed data gives light-tailed data to which the usual mean calculations apply.

Heavy tails and high variability are traditionally acknowledged in distribution systems when calculating reliability metrics by excluding the heavy tails using major event days. However, there is an increasing need for metrics to quantify the risk of extreme weather events in the heavy tails in order to monitor and mitigate resilience.
This quantification currently relies on metrics such as Conditional Value at Risk and other mean values that only work for sufficiently light tails.
We show five cases of distribution utilities in which the heavy tails and huge variations of event costs cause problems with these metrics.
To address this, we propose resilience metrics that can practically describe and take account of the heavy tails.
That is, we address a problem of extreme variability in quantifying the risk of extreme events that occurs in some observed distribution systems' data, and suggest resilience metrics that solve this problem.


\bibliographystyle{IEEEtran}

\end{document}